\documentclass[twocolumn,aps,pre,superscriptaddress,showpacs]{revtex4}
\usepackage[latin9]{inputenc}
\usepackage{amsmath}
\usepackage{amssymb}
\usepackage{graphicx}
\usepackage{esint}

\makeatletter

\usepackage{epsfig}\usepackage{times}

\makeatother

\begin{document}

\title{Signal transmission in a Y-shaped one-way chain}

\author{Xiaoming Liang }
\email{xm-liang@hotmail.com}

\affiliation{School of Physics and Electronic Engineering, Jiangsu Normal University,
Xuzhou 221116, China}

\author{Ming Tang}

\affiliation{Web Sciences Center, University of Electronic Science and Technology
of China, Chengdu 610054, China}

\author{Huaping L\"{u}}

\affiliation{School of Physics and Electronic Engineering, Jiangsu Normal University,
Xuzhou 221116, China}
\begin{abstract}
It has been found that noise plays a key role to improve signal transmission
in a one-way chain of bistable systems {[}Zhang \emph{et al}., Phys.
Rev. E 58, 2952 (1998){]}. We here show that the signal transmission
can be sharply improved without the aid of noise, if the one-way chain
with a single source node is changed with two source nodes becoming
a Y-shaped one-way chain. We further reveal that the enhanced signal
transmission in the Y-shaped one-way chain is regulated by coupling
strength, and that it is robust to noise perturbation and input signal
irregularity. We finally analyze the mechanism of the enhanced signal
transmission by the Y-shaped structure.
\end{abstract}


\maketitle
\textbf{The realization of transmitting weak signals over a long range is essential in engineering.
Stochastic resonance has been proposed as an important mechanism to support such function, where a
weak signal can be transmitted faraway without amplitude attenuation by embedding the nonlinear system
that responsible for transporting the signal in a noisy environment. Subsequently, the nonlinear
systems with complex structures are found to have a higher level of utilizing stochastic resonance
for transmitting signals, as compared to the nonlinear systems with simple and regular structures. However,
the intensity of noise is not easy to be controlled in practice, which reduces the implementation
of stochastic resonance. It is an important question to ask whether there exists a specific structure
by which the signal transmission can be enhanced without the help of noise. For this reason, we here
propose a one-way chain with a Y-shaped structure through modifying the classical one-way chain
model from having a single source node to having two disconnected source nodes. Our results show that
such a slight change in the structure may enable a largely enhanced signal transmission in the one-way
chain. Besides this, the enhanced signal transmission by the Y-shaped structure is much effective
than by stochastic resonance. These findings may contribute to the design of highly efficient
artificial devices.}

\section{INTRODUCTION}

Exploring the relationship between structure and function of real
systems has been improved markedly in recent years, as it has become
clear that the impressive function of real systems is closely related
to their particular structures \cite{Boccaletti:2006,Albert:2002,Newman1:2003,Popovych:2011,Liu1:2012}.
Examples include the high risk of epidemic outbreak in social entities
shared with small-world friendship \cite{Newman2:2003}, the low
threshold of particle condensation in transportation network with
heterogeneous structure \cite{Tang:2008}, and the pathological
brain states accompanied by abnormal anatomical connectivity \cite{Bullmore:2009}.
Signal transmission over long distances is one of the most essential
function in nature, ranging from cell signaling in the nervous system
up to human telecommunication in the engineering \cite{Kumar:2010,McCullen:2007},
but which architecture supports an efficient and robust transmission
is still not fully understood.

Early attempts at exploring the structure-function relationship of signal transmission
were focused on one-way chains \cite{Zhang:1998,Zaikin:2001,Yao:2010,Liu2:2012,Wang:2013}.
In these classical chain models, a node at one side called source node
is responsible for receiving input signals, and then the source node
propagates the signals to its nearest node in single direction, and
so on. It has been reported that a weak signal can be transmitted
along the one-way chain without amplitude damping if the chain is
embedded in noisy environments \cite{Zhang:1998}.
Such noise-improved signal transmission is further observed in complex
networks \cite{Wang:2007,Perc:2007,Perc:2008,Perc:2008b,Perc:2009,Perc:2010,Liu:2008}.
However, the noise-improved signal transmission relies heavily on the proper
intensity of noise which it is hard to be tuned in practice. It is
therefore quite important to seek a specific structure by which the
transmission can be efficiently improved, instead of by the well-tuned
noise.

In this paper, we propose a modified one-way chain model with a Y-shaped
structure and study how such structure affects signal transmission
in the chain. Unlike the classical one-way chain with a single
source node, the Y-shaped one-way chain has two disconnected source
nodes that receive the same input signal. We find that the Y-shaped
one-way chain can maintain long-distance signal transmissions without
amplitude attenuation, no matter
the input signal is periodic or aperiodic. We also
find that the enhanced signal transmission in the Y-shaped one-way
chain is much effective than the noise-improved signal transmission
in the classical one-way chain. These findings imply that even a small
change in the structure might permit a hugely different performance
in signal transmission, offering a good illustration of the relationship
between structure and function.

\begin{figure}
\includegraphics[width=1\linewidth]{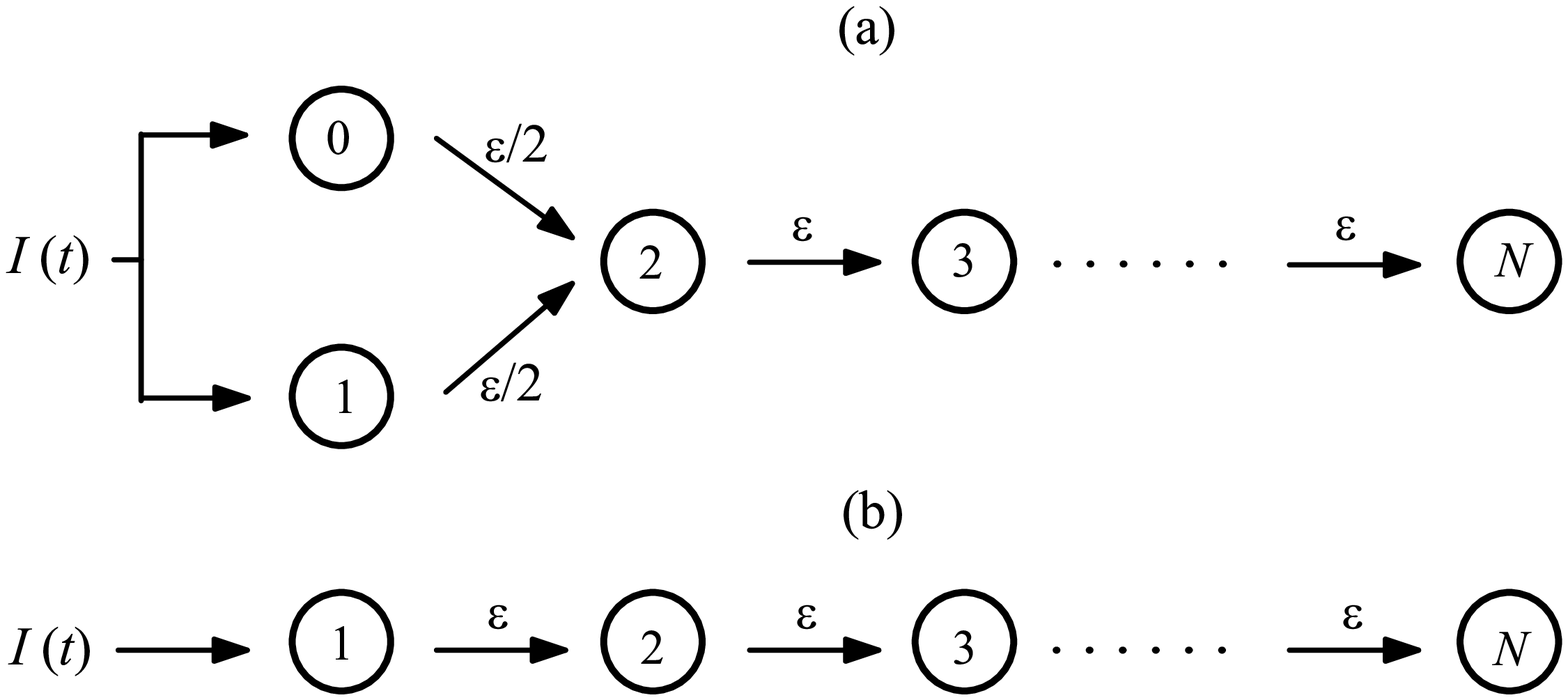} \caption{Architectures of a Y-shaped one-way chain with two disconnected source nodes ($j=0,1$)
to receive an input signal $I(t)$ in (a) and a classical one-way
chain with one source node ($j=1$) to receive the same input signal
$I(t)$ in (b). $\varepsilon$ represents the coupling strength. }

\label{Fig:1}
\end{figure}

\section{MODEL AND METHOD DESCRIPTIONS}

A Y-shaped one-way chain of $N+1$ coupled bistable systems is shown
in Fig. \ref{Fig:1}(a), whose dynamics is described as follows:
\begin{eqnarray}\label{eq:model}
\dot{x}_{j} & = & x_{j}-x_{j}^{3}+I(t),\; j=0,1\nonumber \\
\dot{x}_{2} & = & x_{2}-x_{2}^{3}+\varepsilon\big(\frac{x_{0}+x_{1}}{2}-x_{2}\big),\\
\dot{x}_{j} & = & x_{j}-x_{j}^{3}+\varepsilon(x_{j-1}-x_{j}),\; j=3,\cdots,N\nonumber
\end{eqnarray}
where $\dot{x}_{j}=x_{j}-x_{j}^{3}$ governs the local
dynamics of node $j$, which has two stable fixed points $x_{s}=\pm1$ and one
unstable fixed point $x_{u}=0$, $\varepsilon$ denotes the coupling strength,
and $I(t)$ represents
the input signal receiving by the source nodes ($j=0,1$).
To model weak signal transmissions, $I(t)$ is set as a subthreshold
signal, namely, under such signal, each source node cannot jump between
the two stable fixed points but oscillate around one of them. When
$x_{0}(t)=x_{1}(t)$, the Y-shaped one-way chain of Eq. (\ref{eq:model})
can be viewed as a classical one-way chain with one source
node, see Fig. \ref{Fig:1}(b).

To characterize signal transmission along the chain, we calculate the output of node
$j$ at the frequency $\omega$ of the input signal by \cite{Zaikin:2001,Perc:2008,Liang:2009,Liang:2013}
\begin{equation}
Q_{j}=\left|\frac{\omega}{n\pi}\int_{0}^{\frac{2n\pi}{\omega}}x_{j}(t)e^{i\omega t}dt\right|,\label{eq:indicator}
\end{equation}
where parameter $n$ determines the length of the integration
interval. To achieve a stable result of $Q_j$, a large value of $n=100$ is considered.
Besides, when the input signal is aperiodic or in a noisy environment, $Q_{j}$ is averaged with $100$ realizations.
From Eq. (\ref{eq:indicator}), the signal transmission along the chain is damped if
$Q_{j}>Q_{j+1}$ for $j\geq1$; otherwise, the transmission is enhanced if $Q_{j}\leq Q_{j+1}$
for $j\geq1$. In our discussions, the chain size $N=30$ is used, and the initial
condition $x_{j}(0)$ of each node is randomly selected from the two stable fixed points
$x_{s}=\pm1$. Obviously, the two source nodes display the same dynamical behavior
$x_{0}(t)=x_{1}(t)$ if their initial conditions are identical $x_{0}(0)=x_{1}(0)$, while showing
different dynamical behaves $x_{0}(t)\neq x_{1}(t)$
if their initial conditions are nonidentical $x_{0}(0)\neq x_{1}(0)$. In this regard, Eq. (\ref{eq:model}) with $x_{0}(0)=x_{1}(0)$ and with $x_{0}(0)\neq x_{1}(0)$ represents the classical one-way chain and
Y-shaped one-way chain, respectively.

\begin{figure}
\includegraphics[width=1\linewidth]{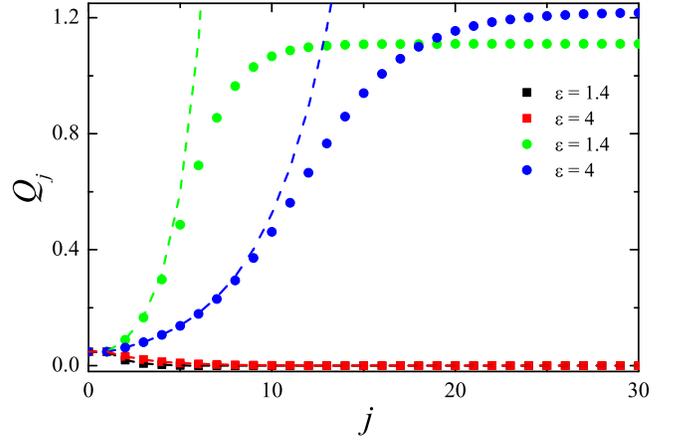} \caption{(Color online) Signal transmissions of Eq. (\ref{eq:model}) with
$x_{0}(0)=x_{1}(0)$ (black and red squares) and with $x_{0}(0)\neq x_{1}(0)$
(green and blue circles) at $\varepsilon=1.4$ and $\varepsilon=4$,
respectively. Dashed lines denote the analytical predictions of Eqs.
(\ref{eq:output-j-1}) and (\ref{eq:output-j-2}). }

\label{Fig:2}
\end{figure}

\section{NUMERICAL RESULTS}
\subsection{Y-shaped structure effect}

A subthreshold periodic signal $I(t)=A\sin(\omega t)$ with $\omega=\pi/5$
and $A=0.1$ is firstly considered. Figure \ref{Fig:2} shows the
transmissions of such signal for two coupling strengths, obtaining
from randomly setting the initial conditions of all the nodes.
It can be observed that $Q_{j}$ always takes only two
distinct responses at each coupling strength: damped transmission
and enhanced transmission. Our numerical results reveal that the former
is achieved at $x_{0}(0)=x_{1}(0)$ while the latter is obtained at
$x_{0}(0)\neq x_{1}(0)$, irrespective of the initial conditions of
the other nodes. Meanwhile, Fig. \ref{Fig:2} shows that the enhanced
signal transmission obtained at $x_{0}(0)\neq x_{1}(0)$ is very sensitive
to the value of coupling strength. When $\varepsilon=1.4$, $Q_{j}$
increases fast and saturates from $j=14$. In contrast, when $\varepsilon=4$,
$Q_{j}$ increases slowly but attains a
higher saturated output after $j\geq25$. Hence, the Y-shaped one-way
chain (at $x_{0}(0)\neq x_{1}(0)$) has a function of enhancing signal
transmission and such function is purely generated by the simple Y-shaped
structure.

\begin{figure}
\includegraphics[width=1\linewidth]{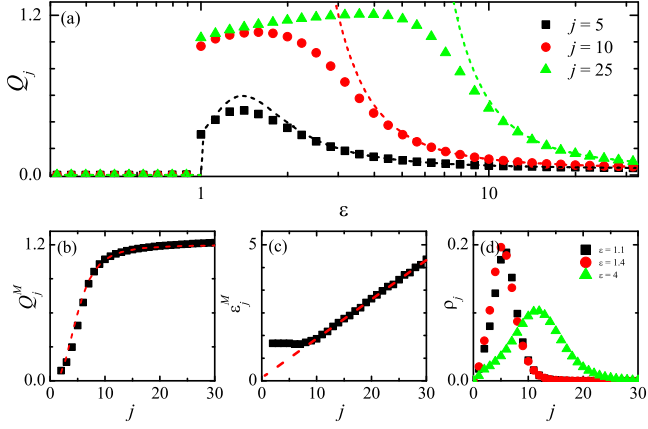} \caption{(Color online) Signal transmission of Eq. (\ref{eq:model}) with $x_{0}(0)\neq x_{1}(0)$.
(a) $Q_{j}$ versus $\varepsilon$ for node $j=5$ (square), $10$
(circle) and $25$ (triangle). Dashed lines denote the analytical
results of Eqs. (\ref{eq:output-j-1}) and (\ref{eq:output-j-2}). (b) The maximum
output $Q_{j}^{M}$ versus $j$ with a fit line $Q_{j}^{M}=1.2j^{3}/(120+j^{3})$.
(c) Optimal $\varepsilon_{j}^{M}$ versus $j$ with a fit line $\varepsilon_{j}^{M}=0.14(1+j)$.
(d) Transmission efficiency $\rho_{j}$ versus $j$ for $\varepsilon=1.1$
(square), $1.4$ (circle), and $4$ (triangle). }

\label{Fig:3}
\end{figure}

The above observations raise two questions: (i) How does the coupling strength impact on the enhanced output
$Q_{j}$ and (ii) which node has the best efficiency of enhancing signal transmission in the Y-shaped one-way
chain? To answer these questions, we compare the dependencies of $Q_{j}$
on $\varepsilon$ between three nodes, see Fig. \ref{Fig:3}(a).
A common feature in this figure is the same critical
coupling strength $\varepsilon_{c}\approx1$ below ($\varepsilon<\varepsilon_{c}$)
or far beyond ($\varepsilon\gg\varepsilon_{c}$) which the output
$Q_{j}\approx0$. In between, the enhanced output $Q_{j}$ emergences
and a maximum output $Q_{j}^{M}$ appears at an optimal coupling
strength $\varepsilon_{j}^{M}$. Moreover,
the intermediate region of $\varepsilon$ with enhanced $Q_j$ is expanded as $j$ increases. During this process,
the values of $Q_{j}^{M}$ and $\varepsilon_{j}^{M}$ are changed accordingly.
As shown in Fig. \ref{Fig:3}(b), $Q_{j}^{M}$ is an increasing function
of $j$ which satisfies $Q_{j}^{M}\approx1.2j^{3}/(120+j^{3})$.
In Fig. \ref{Fig:3}(c), $\varepsilon_{j}^{M}$
seems to be a constant ($\varepsilon_{j}^{M}\approx0.14$) before $j=9$, and then grows with $j$
obeying a linear relationship $\varepsilon_{j}^{M}\approx0.14(1+j)$. Based
on these quantities, we define $\rho_{j}$ to measure the signal transmission
efficiency of node $j$ as
\begin{equation}
\rho_{j}\equiv\frac{Q_{j}-Q_{j-1}}{j-(j-1)}=Q_{j}-Q_{j-1},\; j\geq2\label{eq:rou}
\end{equation}
The results of $\rho_{j}$ for three coupling strengths are given in Fig.~\ref{Fig:3}(d).
It can be observed that $\rho_{j}$ displays a bell-shaped curve at each coupling strength.
In particular, when $\varepsilon=1.1$, the curve of $\rho_{j}$ has a peak at
$j=6$, suggesting that node $j=6$ has the best efficiency of signal
transmission. Interestingly, when $\varepsilon=1.4$, the best transmission
efficiency is gained by node $j=5$ since the peak height at $j=5$ is
higher than at $j=6$. However, when $\varepsilon=4$, the peak of $\rho_{j}$ is shifted to
$j=12$, accompanied by a decline in the peak height. The variations of $\rho_{j}$
indicate that the coupling strength regulates the efficiency of signal
transmission and an intermediate coupling strength enables some node to have a higher
transmission efficiency.

\begin{figure}
\includegraphics[width=1\linewidth]{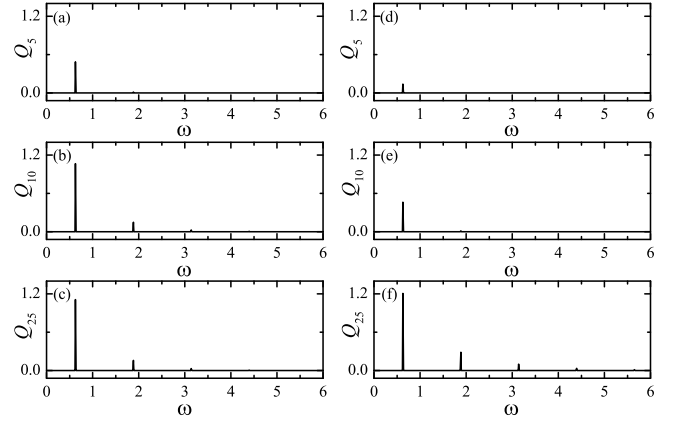} \caption{Spectra of $Q_{j}$. Left panels with
$\varepsilon=1.4$: (a) $j=5$, (b) $j=10$, and (c) $j=25$; Right panels with $\varepsilon=4$: (d) $j=5$,
(e) $j=10$, and (f) $j=25$. Initial condition $x_{0}(0)\neq x_{1}(0)$ is used.}

\label{Fig:4}
\end{figure}

To give a deep insight of the enhanced signal transmission, Fig.~\ref{Fig:4} shows the spectra
of $Q_{j}$ for nodes $j=5$, $10$, and $25$.
When $\varepsilon=1.4$, $Q_{5}$ can be seen as a delta function of $\omega$ which is zero everywhere except at the input frequency $\omega=\pi/5$, where it is a sharp peak, see Fig.~\ref{Fig:4}(a).
Except for the peak at $\omega=\pi/5$, $Q_{10}$ also shows a lower peak
at the harmonic frequency $\omega=3\pi/5$, see Fig.~\ref{Fig:4}(b). Such multiple
peaks can be found for $Q_{25}$, see Fig.~\ref{Fig:4}(c). In addition,
when $\varepsilon=4$, the spectra of $Q_{j}$ are similar to that of $\varepsilon=1.4$, see
Figs.~\ref{Fig:4}(d)-(f). The main difference is that, there are more peaks at other harmonic
frequencies emerge for $Q_{25}$. The emergence of lower peaks at harmonic frequencies means that
the output signal $x_{j}(t)$ is not a pure sine (cosine) wave but a sum of a set of
sine (cosine) waves. However, as the peaks at harmonic frequencies are relatively lower than the peaks at $\omega=\pi/5$, the output $Q_{j}$ at the input frequency
gives a reliable measurement of signal transmission.

\begin{figure}[b!]
\includegraphics[width=1\linewidth]{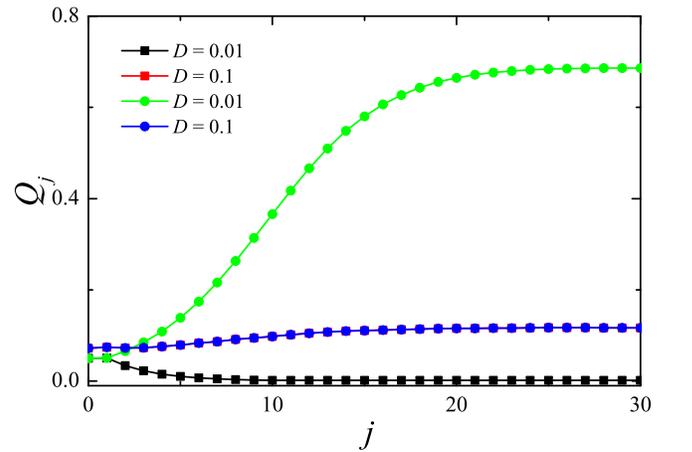} \caption{(Color online) Signal transmission of Eq. (\ref{eq:model}) with
$x_{0}(0)=x_{1}(0)$ (black and red square) and with $x_{0}(0)\neq x_{1}(0)$
(green and blue circles) for $D=0.01$ and $D=0.1$,
respectively. Parameter $\varepsilon=4$ is considered. }
\label{Fig:5}
\end{figure}

\subsection{Robustness to noise}

Since noise is ubiquitous in nature, we examine the robustness of
the enhanced signal transmission in the Y-shaped one-way chain to
external noise perturbation. Hence, each bistable system in Eq. (\ref{eq:model})
becomes noisy, i.e., $\dot{x}_{j}=x_{j}-x_{j}^{3}+\Gamma_{j}(t)$, where
$\Gamma_{j}(t)$ denotes the noise perturbation. We here consider
$\Gamma_{j}(t)$ as the white and spatially uncorrelated noise with
$\langle\Gamma_{j}(t)\rangle=0$ and $\langle\Gamma_{j}(t)\Gamma_{k}(t')\rangle=2D\delta_{jk}\delta(t-t')$,
where parameter $D$ controls the intensity of noise. For a given
coupling strength $\varepsilon=4$, Fig. \ref{Fig:5} shows the transmissions of
the input signal $I(t)=A\sin(\omega t)$ in two noisy environments. In the case of $D=0.01$,
$Q_j$ also displays two distinct responses: damped transmission
at $x_{0}(0)=x_{1}(0)$ and enhanced transmission at $x_{0}(0)\neq x_{1}(0)$.
In the case of $D=0.1$, the
transmission at $x_{0}(0)=x_{1}(0)$ is not damped but slightly enhanced now, which is consistent with the noise-improved signal transmission as observed in \cite{Zhang:1998}. Moreover, such noise-improved transmission at $D=0.1$ displays
the same behavior to the transmission of $x_{0}(0)\neq x_{1}(0)$,
implying that the enhanced signal transmission by the Y-shaped structure is reduced for large
$D$. The phenomenon shown in Fig. \ref{Fig:5} can be understood
as follows. For small $D$, the two source nodes approximate $x_{0}\text{(t)}\approx x_{1}(t)$
if their initial conditions are identical $x_{0}(0)=x_{1}(0)$. Accordingly, Eq. (\ref{eq:model}) consisted of noisy bistable systems can be treated as the classical
one-way chain so that it displays a similar transmission to the case of $D=0$. For large $D$, the noise perturbation is sufficient that it can trigger the source nodes jump between their two stable fixed points. Therefore,
the signal transmission is independent of the initial conditions of
the source nodes, which results in the same transmission between $x_{0}(0)=x_{1}(0)$
and $x_{0}(0)\neq x_{1}(0)$.

\begin{figure}
\includegraphics[width=1\linewidth]{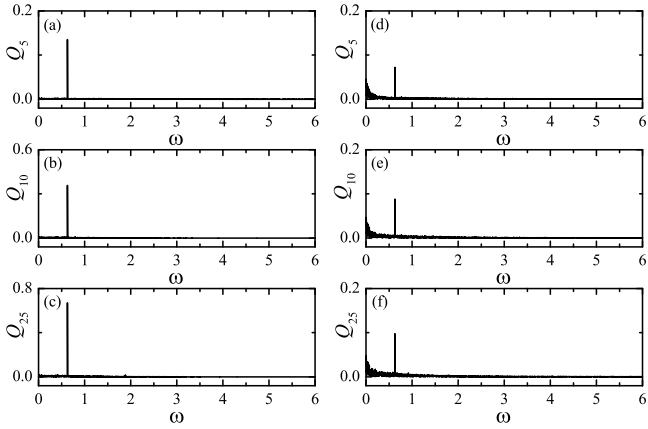} \caption{
Spectra of $Q_{j}$. Left panels with
$D=0.01$: (a) $j=5$, (b) $j=10$, and (c) $j=25$; Right panels with $D=0.1$: (d) $j=5$,
(e) $j=10$, and (f) $j=25$. Parameter $\varepsilon=4$ and initial condition $x_{0}(0)\neq x_{1}(0)$ are used.}
\label{Fig:6}
\end{figure}

Fixed $\varepsilon=4$, we explore the dependency of $Q_{j}$ on $\omega$ for three nodes chosen from Fig.~\ref{Fig:5}.
The results are displayed in Fig.~\ref{Fig:6}. For $D=0.01$, the curve of $Q_{j}$ can be viewed as a delta function
with a sharp peak at the input frequency $\omega=\pi/5$, see Figs.~\ref{Fig:6}(a)-(c).
For $D=0.1$, $Q_{j}$ also resembles a delta function except
small $\omega\approx0$ at which $Q_{j}>0$, see Figs.~\ref{Fig:6}(d)-(f). The common peak at $\omega=\pi/5$ shown in Fig.~\ref{Fig:6} suggests that the input frequency is the main frequency of the output signals and
thus the output $Q_j$ at $\omega=\pi/5$ is the dominant output.

\begin{figure}
\includegraphics[width=1\linewidth]{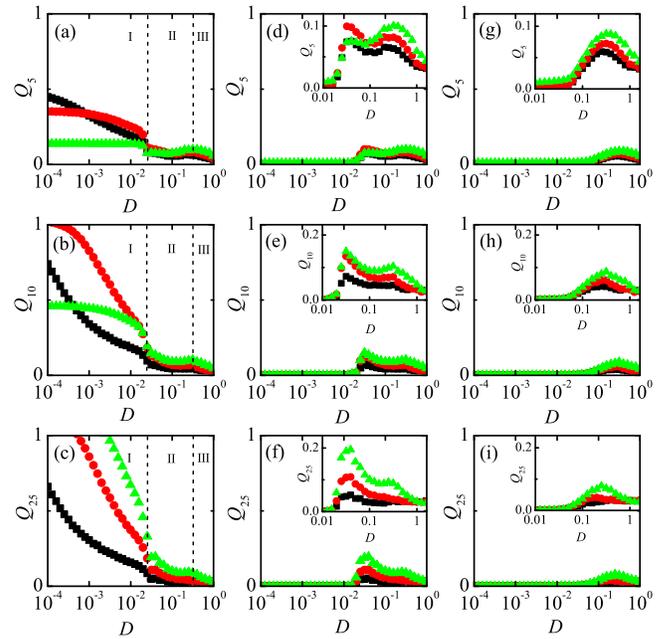} \caption{(Color online) Signal transmission of
Eq. (\ref{eq:model}) at $\varepsilon=1.4$ (square), $2$ (circle), and $4$ (triangle), respectively.
Upper panels for $j=5$: (a) $x_{0}(0)\neq x_{1}(0)$,
(b) $x_{0}(0)=x_{1}(0)$, (c) $x_{0}(0)=x_{1}(0)$ and $\Gamma_{0}(t)=\Gamma_{1}(t)$;
Middle panels for $j=10$: (d) $x_{0}(0)\neq x_{1}(0)$,
(e) $x_{0}(0)=x_{1}(0)$, (f) $x_{0}(0)=x_{1}(0)$ and $\Gamma_{0}(t)=\Gamma_{1}(t)$;
Lower panels for $j=25$: (d) $x_{0}(0)\neq x_{1}(0)$,
(e) $x_{0}(0)=x_{1}(0)$, (f) $x_{0}(0)=x_{1}(0)$ and $\Gamma_{0}(t)=\Gamma_{1}(t)$.
Insets are the enlarged views of signal transmissions.}

\label{Fig:7}
\end{figure}

In addition, the same transmission at large $D$ shown in Fig. \ref{Fig:5} motivates us to
figure out the critical noise intensity at which the signal transmission is irrelevant
to the initial conditions of the source nodes. To this end, we compare
the evolutions of $Q_{j}$ with $D$ between $x_{0}(0)\neq x_{1}(0)$ and $x_{0}(0)= x_{1}(0)$ for several values of $j$ and $\varepsilon$, see Fig. \ref{Fig:7}.
When $x_{0}(0)\neq x_{1}(0)$, $Q_{j}$ decays with $D$ except a slight rise
around $D\approx0.3$, see Figs. \ref{Fig:7}(a)-(c).
When $x_{0}(0)=x_{1}(0)$, $Q_{j}$ suddenly increases from $D\approx0.02$ until attaining a local maximum at $D\approx0.03$,
exhibiting the same performance to the case of $x_{0}(0)\neq x_{1}(0)$ for large $D$, see Figs. \ref{Fig:7}(d)-(f).
When $j$ or $\varepsilon$ varies, the value of $D\approx0.03$ remains constant,
which indicates that $D\approx0.03$ is the critical noise intensity at which the signal transmission in the
Y-shaped one-way chain is not sensitive to the initial conditions of the source nodes.
Besides, Figs. \ref{Fig:7}(d)-(f) (insets) also show that $Q_{j}$ may exhibit two resonant peaks for suitable $\varepsilon$, forming double resonant-like phenomena. Further, Figs. \ref{Fig:7}(g)-(i) (insets) depict the evolutions of $Q_{j}$ for the
classical one-way chain, by setting $x_{0}(0)=x_{1}(0)$ and $\Gamma_{0}(t)=\Gamma_{1}(t)$ in Eq. (\ref{eq:model}). In these figures, $Q_{j}$ shows a resonant-like dependency on $D$ for each pair of $j$ and $\varepsilon$, where the resonant peak is at $D\approx0.3$.
When $D>0.3$,
$Q_{j}$ exhibits a similar evolution to the cases of $x_{0}(0)\neq x_{1}(0)$ and $x_{0}(0)=x_{1}(0)$. This implies that $D\approx0.3$ is another critical noise intensity, above which the difference in signal transmission between the Y-shaped one-way chain and classical one-way chain is small.
Making use of these two critical intensities,
we may divide the signal transmission in the Y-shaped one-way chain
into three regions: region I ($D\leq D_{1}=0.03$), region II ($D_{1}<D<D_{2}=0.3$),
and region III ($D\geq D_{2}$) [see Figs. \ref{Fig:7}(a)-(c)]. Specifically, region I corresponds to
the Y-shaped structure-improved transmission, region II corresponds
the structure-noise-improved transmission, and region III corresponds
the noise-improved transmission, respectively. Among them,
the Y-shaped structure-improved transmission (region
I) is robust to noise perturbation, especially at large $\varepsilon$ since the decay rate is slow.
In addition, the Y-shaped structure-improved transmission is much more
effective than the noise-improved transmission.

\subsection{Robustness to signal irregularity}
\begin{figure}
\includegraphics[width=1\linewidth]{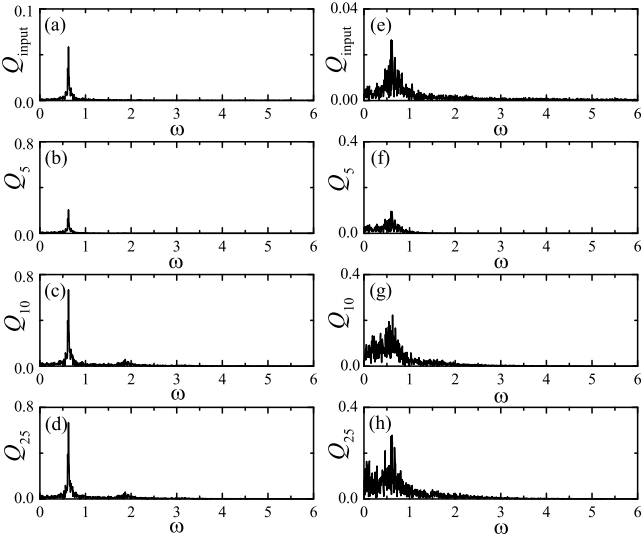} \caption{Spectra of $Q_{j}$.
Left panels with $T=0.01$: (a) $Q_{\text{input}}$, (b) $j=5$, (c) $j=10$, and
(d) $j=25$; Right panels with $T=0.1$: (e) $Q_{\text{input}}$, (f) $j=5$,
(g) $j=10$, and (h) $j=25$. Parameter $\varepsilon=2$ and $x_{0}(0)\neq x_{1}(0)$ are set.}

\label{Fig:8}
\end{figure}

The actual signals are usually irregular ones, it is necessary to check the robustness of the Y-shaped structure-improved transmission to input signal irregularity. Here the irregular input
signal is generated by setting the periodic signal with a
time-varying initial phase $\varphi(t)$, i.e., $I(t)=A\sin\left(\omega t+\varphi(t)\right)$
\cite{Liang:2011}. For simplicity, the initial phase $\varphi(t)$
is set to be varied as a Wiener process. Thus, $\dot{\varphi}(t)=\zeta(t)$
is a Gaussian white noise with $\langle\zeta_{i}(t)\rangle=0$ and
$\langle\zeta(t)\zeta(t')\rangle=2T\delta(t-t')$. When $T>0$, the
periodic signal $I(t)=A\sin(\omega t)$ becomes an aperiodic
signal and its regularity decreases with $T$. To illustrate
it, we show the output spectrum of the aperiodic signal at $T=0.01$
{[}see Fig. \ref{Fig:8}(a){]} and at $T=0.1$ {[}see Fig. \ref{Fig:8}(e){]},
respectively. In both spectra, there is a highest peak at $\omega\approx\pi/5$,
where the peak height is lower and peak width
is wider at $T=0.1$, demonstrating that the regularity of the
aperiodic signal is decreased with $T$. We next investigate whether these two aperiodic
signals can be effectively transmitted in the Y-shaped one-way chain.
Fixing $\varepsilon=2$, Figs. \ref{Fig:8}(b)-(d) depict the output
spectra for three nodes $j=5$, $10$, and $25$ at $T=0.01$.
It is obvious that each output spectrum can be
considered as an enlarged version of Fig. \ref{Fig:8}(a),
where the output spectra of $Q_{10}$ and $Q_{25}$ show larger enlarged
ratios than that of $Q_{5}$. Similarly, such enlarged versions can also
be observed in Figs. \ref{Fig:8}(e)-(h) for the case of $T=0.1$. Comparing with that of $T=0.01$, the enlarged ratio
and fidelity are reduced at $T=0.1$. From these observations, it can be concluded that
the Y-shaped structure-improved transmission works well for irregular
signals.

\section{ANALYTICAL RESULTS}

We now analyze the underlying mechanism of the Y-shaped structure-improved
signal transmission. To avoid the effect of noise, we only discuss
Eq. (\ref{eq:model}) subjected to a periodic input signal ($T=0$) in absence
of noise ($D=0$). Because the input signal $I(t)=A\sin(\omega t)$ is
subthreshold, the source nodes oscillate with small amplitudes around the stable
fixed points, their solutions can be approximately obtained as \cite{Liang:2009,Acebron:2007,Liang:2012}
\begin{equation}\label{eq:input nodes}
x_{j}(t)\approx x_j(0)+A_1\sin\left(\omega t+\varphi_{1}\right),\: j=0,1
\end{equation}
where $x_j(0)=\pm1$ depending on the initial condition, $A_1={A}/{\sqrt{\omega^{2}+4}}$,
and $\varphi_{1}$ denotes some phase shift.

\subsection{Case 1: Two source nodes with the same initial condition}
When $x_{0}(0)=x_{1}(0)$, the dynamical equation of node $j=2$ becomes
\begin{equation}\label{eq:second layer-1}
\dot{x}_{2}=(1-\varepsilon)x_{2}-x_{2}^{3}+\varepsilon x_1(0)+\varepsilon A_1\sin\left(\omega t+\varphi_{1}\right).
\end{equation}
Without the periodic signal $\varepsilon A_1\sin\left(\omega t+\varphi_{1}\right)$, $x_2$
has three fixed points for $\varepsilon\leq1/4$: $x_1(0)$ and $-1/2\pm\sqrt{1-4\varepsilon}/2$ in which $x_1(0)$ and $-1/2-\sqrt{1-4\varepsilon}/2$ are stable fixed points while $-1/2+\sqrt{1-4\varepsilon}/2$ is unstable; for $\varepsilon>{1}/{4}$, $x_2$ has one stable fixed point $x_1(0)$. When $\varepsilon$ is not great, the signal $\varepsilon A_1\sin\left(\omega t+\varphi_{1}\right)$ is subthreshold, the solutions of the node $j=2$ approximate
\begin{equation}\label{eq:s-1 of n=2-1}
{x}_{2}(t)\approx x_1(0)+\frac{\varepsilon A_1}{\sqrt{\omega^{2}+(2+\varepsilon)^{2}}}\sin\left(\omega t+\varphi_{2}\right)\nonumber
\end{equation}
and
\begin{equation}\label{eq:s-2 of n=2-1}
{x}_{2}(t)\approx \frac{-1-\sqrt{1-4\varepsilon}}{2}+\frac{\varepsilon A_1}{\sqrt{\omega^{2}+(\frac{1}{4}-\varepsilon)^{2}}}\sin\left(\omega t+\varphi_{2}\right),\nonumber
\end{equation}
where $\varphi_{2}$ is some phase shift. When $\varepsilon\approx1/4$, the latter solution indicates a larger oscillation around $-1/2-\sqrt{1-4\varepsilon}/2$ than the former around $x_1(0)$. However, the stability of the fixed point $-1/2-\sqrt{1-4\varepsilon}/2$ decreases as $\varepsilon$ approaches $1/4$, the large oscillation is thus unsustainable and it will move to the vicinity of $x_1(0)$, leading to a small oscillation governed by the former solution.
Inserting $x_2(t)$ into the equation of node $j=3$, we can obtain the stable fixed points of the node $j=3$ as well as the
subsequent nodes by repeatedly using the same method.
We find that these nodes possess the same stable fixed point $1$ or $-1$, depending on $x_1(0)=1$ or $x_1(0)=-1$. In this way, the dynamical equation of node $j\geq3$ can be written as
\begin{equation}\label{eq:eq-j-1}
\dot{{x}}_{j}\approx (1-\varepsilon)x_j-{x_j}^{3}+\varepsilon x_1(0)+\varepsilon A_{j-1}\sin(\omega t+\varphi_{j-1}),
\end{equation}
where $\varepsilon A_{j-1}\sin(\omega t+\varphi_{j-1})$ denotes the signal from node $j-1$ and $\varphi_{j-1}$ represents some phase shift. When the signal $\varepsilon A_{j-1}\sin(\omega t+\varphi_{j-1})$ is subthreshold, the solution of Eq. (\ref{eq:eq-j-1}) approximately satisfies
\begin{eqnarray}\label{eq:solution-j-1}
{x}_{j}(t)&\approx& x_1(0)+\frac{\varepsilon A_{j-1}}{\sqrt{\omega^{2}+(2+\varepsilon)^{2}}}\sin\left(\omega t+\varphi_{j}\right)\approx x_1(0)\nonumber\\
& &+\left(\frac{\varepsilon}{\sqrt{\omega^{2}+(2+\varepsilon)^{2}}}\right)^{j-1}A_{j-1}\sin\left(\omega t+\varphi_{j}\right)
\end{eqnarray}
with some phase shift $\varphi_{j}$. Inserting this solution into Eq. (\ref{eq:indicator}),
the output $Q_{j}$ is given by
\begin{equation}\label{eq:output-j-1}
Q_{j}\approx A_1\left(\frac{\varepsilon}{\sqrt{\omega^{2}+(2+\varepsilon)^{2}}}\right)^{j-1}.
\end{equation}
Eq. (\ref{eq:output-j-1}) satisfies the condition $Q_{j}>Q_{j+1}$ for $j\geq1$,
thereby supporting the damped transmission of Eq. (\ref{eq:model}) at $x_{0}(0)=x_{1}(0)$.

\begin{figure}
\includegraphics[width=1\linewidth]{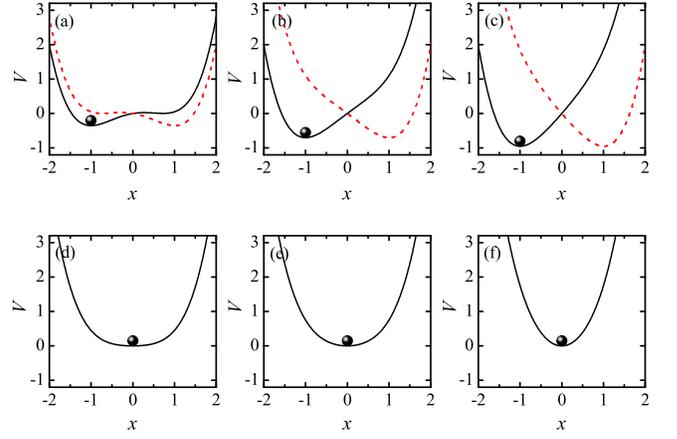} \caption{(Color online) Potential $V(x)$ for diffent $\varepsilon$.
Upper panels for $V(x)=-(1-\varepsilon)\frac{x^2}{2}+\frac{x^4}{4}-\varepsilon x_1(0) x$: (a) $\varepsilon=0.2$, (b) $\varepsilon=0.9$, and (c) $\varepsilon=1.4$. Solid lines correspond to $x_1(0)=1$ and the dashed
lines correspond to $x_1(0)=-1$. Lower panels for $V(x)=-(1-\varepsilon)\frac{x^2}{2}+\frac{x^4}{4}$: (d) $\varepsilon=1.4$, (e) $\varepsilon=2$, and (f) $\varepsilon=4$. }

\label{Fig:9}
\end{figure}

On the other hand, the damped transmission at $x_{0}(0)=x_{1}(0)$ can be explained by the overdamped motion
of a particle in a potential and periodic forcing \cite{Gammaitoni:1998}.
For this reason, the potential in Eq. (\ref{eq:eq-j-1}) is $V(x)=-(1-\varepsilon)x^2/2+x^4/4+\varepsilon x_1(0)x$
and the periodic forcing is $\varepsilon A_{j-1}\sin(\omega t+\varphi_{j-1})$.
When $\varepsilon>0$, $V(x)$ is an asymmetrical potential and its asymmetry is determined
by the value of $\varepsilon$. For illustration, Figs. \ref{Fig:9}(a)-(c) display the potential $V(x)$ for
$\varepsilon=0.2$, $0.9$ and $1.4$. When $\varepsilon=0.2$, $V(x)$ has two wells, where the well located at $x=1$ (or $x=-1$) is deeper than the other one at $x\approx0.7$ (or $x\approx-0.7$), see Fig. \ref{Fig:9}(a). This indicates that the large oscillations around $x=1$ (or $x=-1$) are more stable. When $\varepsilon$ is increased to $0.9$, $V(x)$ turns into an V-shaped potential with a single well at $x=1$ (or $x=-1$), see Fig. \ref{Fig:9}(b). As shown in Fig. \ref{Fig:9}(c), further increasing $\varepsilon$ to $1.4$ will result in a more steep V-shaped potential.
Clearly, under the same forcing of $\varepsilon A_{j-1}\sin(\omega t+\varphi_{j-1}$), the asymmetrical potential $V(x)$ of $\varepsilon=0.2$ allows the particle to generate a relatively large oscillation inside it in contrast to the potentials of $\varepsilon=0.9$ and $\varepsilon=1.4$. However, as $\varepsilon A_{j-1}\sin(\omega t+\varphi_{j-1}$) is weak and the motion is overdamped, the oscillation around $x=1$ (or $x=-1$) gets even smaller ($A_{j}<A_{j-1}$). Altogether, the transmission of Eq. (\ref{eq:model}) decreases with $j$ and $\varepsilon$ when $x_{0}(0)=x_{1}(0)$.

\subsection{Case 2: Two source nodes with different initial conditions}
When $x_{0}(0)\neq x_{1}(0)$, Eq. (\ref{eq:second layer-1}) can be rewritten as
\begin{equation}\label{eq:second layer-2}
\dot{x}_{2}=(1-\varepsilon)x_{2}-x_{2}^{3}+\varepsilon A_1\sin\left(\omega t+\varphi_{1}\right).
\end{equation}
Without the periodic signal $\varepsilon A_1\sin\left(\omega t+\varphi_{1}\right)$,
$x_{2}$ has two stable fixed points $\pm\sqrt{1-\varepsilon}$ for
$\varepsilon<1$ and has one stable fixed
point $0$ for $\varepsilon\geq1$. For a subthreshold signal $\varepsilon A_1\sin\left(\omega t+\varphi_{1}\right)$, the solutions of $x_2(t)$ approximate
\begin{equation}\label{eq:s-1 of n=2-2}
{x}_{2}(t)\approx\pm\sqrt{1-\varepsilon}+\frac{\varepsilon A_1}{\sqrt{\omega^{2}+4(1-\varepsilon)^{2}}}\sin\left(\omega t+\varphi'_{2}\right)\nonumber
\end{equation}
and
\begin{equation}\label{eq:s-1 of n=2-2}
{x}_{2}(t)\approx \frac{\varepsilon A_1}{\sqrt{\omega^{2}+(1-\varepsilon)^{2}}}\sin\left(\omega t+\varphi'_{2}\right),\nonumber
\end{equation}
where $\varphi'_{2}$ is some phase shift. Based on the solutions of ${x}_{2}(t)$,
we can acquire the stable fixed points of the subsequent nodes $j\geq3$. We find that the stable fixed points of these nodes are $\pm1$ for $\varepsilon<1$ and $0$ for $\varepsilon\geq1$. In the former case, the dynamical equation of node $j\geq3$ can be rewritten as
\begin{equation}\label{eq:eq-j-2-a}
\dot{{x}}_{j}\approx (1-\varepsilon)x_j-{x_j}^{3}+\varepsilon x_1(0)+\varepsilon A_{j-1}\sin(\omega t+\varphi'_{j-1}),
\end{equation}
where $\varphi'_{j-1}$ is some phase shift. Eq. (\ref{eq:eq-j-2-a}) has the same form as Eq. (\ref{eq:eq-j-1}), so their solutions and the corresponding outputs are similar. This means the signal transmission is damped for $\varepsilon<1$ no matter the initial condition is $x_{0}(0)=x_{1}(0)$ or $x_{0}(0)\neq x_{1}(0)$. In the latter case, i.e., $\varepsilon\geq1$,
the dynamics equation of node $j\geq3$ is
\begin{equation}\label{eq:eq-j-2-b}
\dot{{x}}_{j}\approx (1-\varepsilon)x_j-{x_j}^{3}+\varepsilon A_{j-1}\sin(\omega t+\varphi'_{j-1}).
\end{equation}
Its solution is
\begin{equation}\label{eq:solution-j-2}
{x}_{j}(t)\approx\left(\frac{\varepsilon}{\sqrt{\omega^{2}+(1-\varepsilon)^{2}}}\right)^{j-1}A_{1}\sin\left(\omega t+\varphi'_{j}\right),
\end{equation}
where $\varphi'_{j}$ is some phase shift. Inserting Eq. (\ref{eq:solution-j-2}) into Eq. (\ref{eq:indicator}), the output is
\begin{equation}\label{eq:output-j-2}
Q_{j}\approx A_1\left(\frac{\varepsilon}{\sqrt{\omega^{2}+(1-\varepsilon)^{2}}}\right)^{j-1}.
\end{equation}
Eq. (\ref{eq:output-j-2}) satisfies the condition $Q_{j}\leq Q_{j+1}$
for $j\geq1$, which coincides with the enhanced signal transmissions at $x_{0}(0)\neq x_{1}(0)$. In Fig. \ref{Fig:2}
and Fig. \ref{Fig:3}(a), we compare the analytical results of Eqs.
(\ref{eq:output-j-1}) and (\ref{eq:output-j-2}) with the numerical
results and find a good agreement between them for small
$j$. The reason is that, the above analyses are based on the perturbation
theory, i.e., assuming $x_{j}$ oscillates around the stable fixed point with a small amplitude. Because the oscillation of $x_{j}$ is weak for small $j$, the theory gives a better approximation to $x_{j}$
as well as $Q_{j}$. In addition, from Eq. (\ref{eq:output-j-2}), the optimal $\varepsilon_{j}^{M}$
can be derived as $\varepsilon_{j}^{M}=1+\omega^{2}\approx1.4$, which
fits well with the numerical results ($j\leq9$) shown in Fig. \ref{Fig:3}(b).

Analogously, the enhanced signal transmission at $x_{0}(0)\neq x_{1}(0)$ and $\varepsilon\geq1$ can also be understood by the overdamped motion of a
particle in a potential and periodic forcing. As shown in Eq. (\ref{eq:eq-j-2-b}), the periodic forcing is $\varepsilon A_{j-1}\sin\left(\omega t+\varphi_{j-1}\right)$, and the
potential is $V(x)=-(1-\varepsilon){x^2}/{2}+{x^4}/{4}$ which is a symmetrical function
with a minimum at $x=0$. In Fig. \ref{Fig:9}(d), the potential $V(x)$ for $\varepsilon=1.4$ is plotted. It is a U-shaped curve with a flat bottom, which is quite different from the V-shaped well shown in Fig. \ref{Fig:9}(c). In addition, Fig. \ref{Fig:9}(e) plots the potential $V(x)$ for $\varepsilon=2$. It can be seen that the bottom of the U-shaped $V(x)$ becomes narrow and such narrow U-shaped potential transforms into a V-shaped curve as $\varepsilon=4$, see Fig. \ref{Fig:9}(f). In contrast, the U-shaped potential $V(x)$ can permit the particle to gain a wider oscillation inside it than the V-shaped potentials. This explains why the signal transmission is largely enhanced at $x_{0}(0)\neq x_{1}(0)$ and $\varepsilon=1.4$.

\subsection{Mechanisms of resonant-like phenomena}
\begin{figure}
\includegraphics[width=1\linewidth]{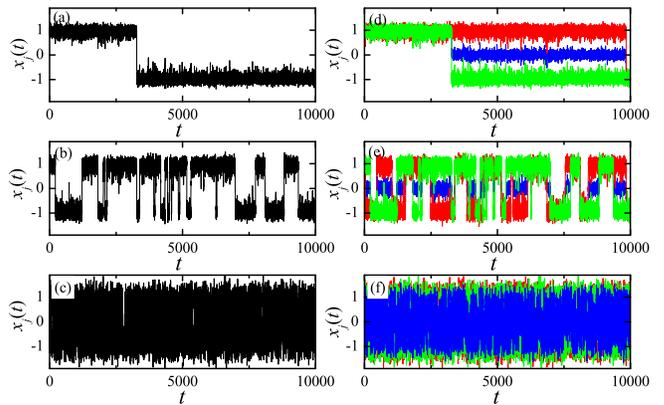} \caption{(Color online) Time series $x_j(t)$ of the source node(s).
Left panels for one source node: (a) $D=0.03$, (b) $D=0.05$, and (c) $D=0.3$;
Right panels for two source nodes with given the same initial condition $x_0(0)=x_1(1)$: (d) $D=0.03$, (e) $D=0.05$, and (f) $D=0.3$. Red and green lines denote $x_j(t)$ of the two source nodes,
blue lines denote the their collective dynamics $(x_0(t)+x_1(t))/2$.}
\label{Fig:10}
\end{figure}

We finally analyze the mechanism of the resonant-like phenomena shown in Fig. \ref{Fig:7}.
Firstly, we explain the single resonant-like dependency for the classical one-way
chain with one source node, i.e., $x_{0}(0)=x_{1}(0)$ and $\Gamma_{0}(t)=\Gamma_{1}(t)$ are set in
Eq. (\ref{eq:model}).
When $D=0$, the oscillation of the source node is small, restricting in one of the two stable fixed points.
When $D$ is increased to $D=0.03$, the oscillation of the source node can jump to the other stable fixed point by noise perturbation, see Fig. \ref{Fig:10}(a). Because the perturbations are not sufficient, the jumping rate is small and the oscillation may stay there for a long time until the next jumping. Thus the oscillation of the source node is still small at $D=0.03$. Continue increasing $D$ to $D=0.05$,
the jumping rate between the two stable fixed points is obviously improved,
which increases the oscillation amplitude, see Fig. \ref{Fig:10}(b). When $D=0.3$, the jumping rate is sharply improved, so the oscillation is no longer centered on the stable fixed points $x_s=\pm1$ but on $x_u=0$, see Fig. \ref{Fig:10}(c). However, further increase in $D$ will increase the randomness of the oscillation (not shown here).
Considering all of these factors, the source node can only generate a large output at $D=0.3$, showing a resonant peak over there. Through one-way coupling, the output of the source node will propagate to the subsequent nodes ($j\geq2$), which results in the stochastic resonance phenomena as observed in Figs. \ref{Fig:7}(g)-(i).

Secondly, we explain the double resonant-like dependency for the Y-shaped one-way chain with $x_{0}(0)=x_{1}(0)$. As mentioned above, when $D=0.03$, there is a small probability that a single source node may jump to the other stable fixed point, remaining there for a long time until it jumps back to the initial fixed point. In this way, the two source nodes in the Y-shaped one-way chain may occasionally oscillate in different stable fixed points for long time intervals, although given the same initial condition $x_{0}(0)=x_{1}(0)$, see Fig. \ref{Fig:10}(d).
Considering that the signal transmission is largely enhanced if
the two source nodes oscillate in different stable fixed points [see Sec. IV B],
the transmission in the Y-shaped one-way chain will be sometimes largely enhanced at $D=0.03$.
By increasing $D$ to $D=0.05$, the time intervals for the two source nodes simultaneously oscillating at different fixed points reduce dramatically [see Fig. \ref{Fig:10}(e)], indicating a decrease in signal transmission.
These are the reasons why $Q_j$ shows the first local peak at $D=0.03$. When $D$ is increased to $D=0.3$,
there is no obvious interval between two continuous jumps, see Fig. \ref{Fig:10}(f).
During this process, the collective behavior of the two source nodes $(x_0(t)+x_1(t))/2$ is analogous to the individual $x_0(t)$ or $x_1(t)$, i.e., the two source nodes can be seen as a single one. This analogy in dynamics implies that the Y-shaped one-way chain shows a similar signal transmission to that of the classical one-way chain for large $D$. As a result, the signal transmission in the Y-shaped one-way chain is also largely enhanced at $D=0.3$, resulting in the second peak over there. Obviously, both the single and double resonant-like dependencies are the
stochastic resonance phenomena, since the signal transmissions are improved by noise. However, as the specific Y-shaped structure allows the two source nodes oscillate in distinct fixed points for small noise, we thus refer the enhanced signal transmission in the region $0.03<D<0.3$ as structure-noise-improved transmission [see Figs. \ref{Fig:7}(d)-(f)].

\section{SUMMARY}
In conclusions, we have studied the signal transmission in a Y-shaped
one-way chain and found an extraordinarily of such specific structure
to improve signal transmission. We have also studied the robustness
of the Y-shaped structure-improved transmission to the noise perturbation
and input signal regularity. We hope our findings may contribute to
understand the structure-function relationship of real systems and
be useful to design highly efficient artificial devices, such as switchers
and amplifiers.

\section*{ACKNOWLEDGMENTS}

X.L. was supported by the NNSF of China
under Grant No. 11305078, the Research Fund of Jiangsu Normal University
under Grant No. 12XLR028, and the Priority
Academic Program Development of Jiangsu Higher Education
Institutions (PAPD). M.T. was supported by the NNSF of China
under Grant No. 11105025. H.L. was supported by the NNSF of China
under Grant No. 11175150.


\begin{references}
\bibitem{Boccaletti:2006} S. Boccaletti, V. Latora, Y. Moreno, M.
Chavez, and D.-U Hwang, Phys. Rep. \textbf{424}, 175 (2006).

\bibitem{Albert:2002} R. Albert and A. Barab\'{a}i, Rev. Mod. Phys.
\textbf{74}, 47 (2002).

\bibitem{Newman1:2003} M. E. J. Newman, SIAM Review \textbf{45},
167 (2003).

\bibitem{Popovych:2011} O. V. Popovych, S. Yanchuk, and P. A. Tass,
Phys. Rev. Lett. \textbf{107}, 228102 (2011).

\bibitem{Liu1:2012} Z. Liu, B. Li, and Y.-C. Lai, EPL \textbf{98},
20005 (2012).

\bibitem{Newman2:2003} M. E. J. Newman, Phys. Rev. E \textbf{61},
5678 (2000).

\bibitem{Tang:2008}M. Tang and Z. Liu, Physica A \textbf{387}, 1361
(2008).

\bibitem{Bullmore:2009} E. Bullmore and O. Sporns, Nat. Rev. Neurosci.
\textbf{10}, 186 (2009).

\bibitem{Kumar:2010} A. Kumar, S. Rotter, and A. Aertsen, Nat. Rev.
Neurosci. \textbf{11}, 615 (2010).

\bibitem{McCullen:2007} N. J. McCullen, T. Mullin, and M. Golubitsky,
Phys. Rev. Lett. \textbf{98}, 254101 (2007).

\bibitem{Zhang:1998} Y. Zhang, G. Hu, and L. Gammaitoni, Phys. Rev.
E \textbf{58}, 2952 (1998).

\bibitem{Zaikin:2001} A. A. Zaikin, J. Garc\'{\i}a-Ojalvo, L. Schimansky-Geier,
and J. Kurths, Phys. Rev. Lett. \textbf{88}, 010601 (2001).

\bibitem{Yao:2010} C. Yao and M. Zhan, Phys. Rev. E \textbf{81},
061129 (2010).

\bibitem{Liu2:2012} Z. Liu, EPL \textbf{100}, 60002 (2012).

\bibitem{Wang:2013} J. Wang and Z. Liu, EPL \textbf{102}, 10003 (2013).

\bibitem{Wang:2007} Q. Wang, Q. Lu, and G. Chen, EPL \textbf{77},
10004 (2007).

\bibitem{Perc:2007} M. Perc, Phys. Rev. E \textbf{76}, 066203 (2007).

\bibitem{Perc:2008} M. Perc, Phys. Rev. E \textbf{78}, 036105 (2008).

\bibitem{Perc:2008b} M. Perc and M. Gosak, New J. Phys. \textbf{10}, 053008 (2008).

\bibitem{Perc:2009} M. Ozer, M. Perc, and M. Uzuntarla, Phys. Lett. A \textbf{373}, 964 (2009).

\bibitem{Perc:2010} X. Sun, M. Perc, Q. Lu, and J. Kurths, Chaos \textbf{20}, 033116 (2010).

\bibitem{Liu:2008} Z. Liu and T. Munakata, Phys. Rev. E \textbf{78},
046111 (2008).

\bibitem{Liang:2009} X. Liang, Z. Liu, and B. Li, Phys. Rev. E \textbf{80},
046102 (2009).

\bibitem{Liang:2011} X. Liang, L. Zhao, and Z. Liu, Phys. Rev. E
\textbf{84}, 031916 (2011).

\bibitem{Acebron:2007} J. A. Acebr\'{o}, S. Lozano, and A. Arenas, Phys.
Rev. Lett. \textbf{99}, 128701 (2007).

\bibitem{Liang:2012} X. Liang, L. Zhao, and Z. Liu, Chaos \textbf{22},
023128 (2012).

\bibitem{Liang:2013} X. Liang, S. Yanchuk, and L. Zhao, Phys. Rev. E
\textbf{88}, 012910 (2013).

\bibitem{Gammaitoni:1998} L. Gammaitoni, P. H\"{a}nggi, P. Jung, and F. Marchesoni,
Rev. Mod. Phys. \textbf{70}, 223 (1998).



\end{references}
\end{document}